\documentclass{moriond}

\usepackage{amsmath,amsfonts,amssymb,epsfig}
\usepackage{amsthm}
\usepackage{latexsym}
\usepackage[utf8]{inputenc}

\def\be{\begin{equation}}
\def\ee{\end{equation}}
\def\bea{\begin{eqnarray}}
\def\eea{\end{eqnarray}}

\def\udark{U(1)_D}

\def\aem{\alpha_{\rm em}}
\def\azp{\alpha_{D}}

\def\mzp{M_{V}}
\def\ms{M_{S}}

\def\eps{\varepsilon}

\newcommand{\ap}{V}

\newcommand{\Photo}{}

\begin{document}
	\vspace*{1cm}
\title{Light dark sector at colliders and fixed target experiments}

\author{Luc Darm\'e,$^{1}$\footnote{Based on a talk given by LD at the 53\textsuperscript{rd} Rencontres de Moriond session devoted to QCD and high energy interactions.} Soumya Rao,$^{1}$ Leszek Roszkowski$^{1,2}$ }

\address{$^{1}$National Centre for Nuclear Research (NCBJ), 69 ul. Ho\.za, 00-681 WARSZAWA, POLAND \\
$^{2}$ Consortium for Fundamental Physics, Department of Physics and Astronomy,
	University of Sheffield, Sheffield S3 7RH, United Kingdom}

\maketitle\abstracts{	\vspace*{0.5cm}
Minimal scenarios with light (sub-GeV) thermal dark matter are usually accompanied by a correspondingly light ``dark sector". Taking as an example a simple fermionic dark matter model, we will show that the presence of the dark sector plays a key role in constraining such scenarios at accelerators experiments. The effect of including a dark Higgs boson in the light spectrum is in particular investigated.	\vspace*{0.5cm}}

\section{Light fermionic dark matter}

Among the numerous solutions of the Dark Matter (DM) puzzle, the ones relying on the thermal freeze-out mechanism have been long considered among the most attractive. 
The prime example of a dark matter scenario based on this mechanism, the Weakly Interactive Massive Particle (WIMP), has been intensely searched for in the past decades. However, the  absence  so far of any uncontroversial signals from colliders, direct and indirect detection experiments has led the community to explore simultaneously numerous other possibilities. Among them the so-called thermal hidden sector scenarios are particularly attractive in that they closely resembled the WIMP while relaxing various experimental bounds and allowing for DM candidates with lighter, sub-GeV masses. The trade-off is that the DM annihilation needs to be generally mediated by a new light field since weak interaction-driven processes are typically too suppressed at this scale. 

In this work we will focus on the case  where the mediator is a new massive
gauge boson $\ap$,~\footnote{The case of a light scalar is already strongly constrained by heavy meson decays.~\cite{Dolan:2014ska}} called henceforth the dark photon, corresponding to an abelian gauge group
$\udark$ spontaneously broken by the Vacuum Expectation Value (VEV) $v_S$ of a dark Higgs boson $S$. We will focus on the case of fermionic dark matter (the case of complex scalar DM is also viable even though more constrained~\cite{Darme:2017glc}).

In order to keep the additional $\udark$ anomaly-free, one minimally needs to introduce two Majorana fermions with opposite charges. We thus consider a Dirac fermion $\chi = (\chi_L,\bar{\chi}_R)$ with $\udark$-charge $-1$ and choose the dark Higgs boson $U(1)_D$ charge to be $+2$ to allow for Yukawa couplings with the left and right-handed components. Writing $A_{\mu\nu}$ the electromagnetic field strength and $F^{\prime}_{\mu\nu}$ the $\udark$ one, the Lagrangian for the dark Higgs boson and dark photon contains:
\begin{eqnarray}
\mathcal{L}_{\ap} &=~
-\frac{1}{2} \varepsilon A_{\mu\nu}F^{\prime\mu\nu} + (D^\mu S )^*(D_\mu S)+ \mu_S^2 |S|^2 - \frac{\lambda_S}{2} |S|^4  \ , 
\label{lma}
\end{eqnarray}
where we have included a kinetic mixing term $\varepsilon$ and have neglected the quartic mixing between the dark Higgs boson and the Standard Model (SM) Higgs.~\cite{Darme:2017glc} The DM part of the Lagrangian is
\begin{eqnarray}
\mathcal{L}^{\rm DM}&=\bar{\chi}\left( i \gamma_\mu D^\mu-m_{\chi}\right)\chi +
y_{SL} S \bar{\chi}^c P_L \chi + y_{SR} S \bar{\chi}^c P_R  \chi + \textrm{ h.c.} \ .
\end{eqnarray}
After the  dark Higgs boson acquires its VEV, the dark matter mass matrix becomes
\begin{equation}
\begin{array}{c} \\ M_\chi = \\ \end{array}
\begin{array}{cc}
& \begin{array}{cc}  \hspace{-10pt} {}_{\chi_L} &  \hspace{30pt} _{\chi_R} \end{array} \\[0.3em]
\begin{array}{c}
_{\bar{\chi}_L} \\
_{\bar{\chi}_R} 
\end{array}  \hspace{-12pt}& \left( \begin{array}{cc}
\sqrt{2} v_S y_{SL} & m_\chi \\
m_\chi & \sqrt{2} v_S y_{SR} \
\end{array} \right) 
\end{array}\, ,
\end{equation}
and is diagonalised by introducing the two Majorana eigenstates $(\chi_1, \chi_2)$. 
 Choosing to keep the rotation matrices $U_\chi$ real then implies that $\chi_1$ has a negative mass~\footnote{Note the different convention from the standard reference.~\cite{TuckerSmith:2001hy}} when the off-diagonal contribution dominates but allows to write the $\udark$ current in the simple form:
\begin{eqnarray}
 \mathcal{J}_D^\mu = \frac{1}{2} \bar{\chi}_i \gamma^\mu \gamma^5 \chi_j \left( U_\chi^{i2}U_\chi^{j2}  - U_\chi^{i1}U_\chi^{j1}\right) \ . 
 \label{eq:darkcurrent}
\end{eqnarray}
In the rest of this work, we will note $M_{\chi_1}$ (respectively $M_{\chi_2}$) the absolute mass of the lightest (heaviest) eigenstate.

Overall the ``dark sector'' hence contains: a Majorana dark matter $\chi_1$, a heavier Majorana state $\chi_2$, a dark photon $V$ and a dark Higgs boson $S$.
In the rest of this work, we will be interested in keeping generic values for the Yukawa couplings, departing from the standard (technically natural) choice $y_{SL} = y_{SR} \ll 1$.
This very simple model leads to the correct relic density in three  regimes with starkly different phenomenology:~\cite{Darme:2017glc}
the pseudo-Dirac regime (iDM), the Majorana DM
case and finally a secluded regime where $\chi_1 \chi_1 \rightarrow SS$ dominates (even though the latter suffers from strong bounds from Big Bang Nucleosynthesis (BBN)-related observables).~\footnote{In this proceedings, we will keep $M_{\chi_1} < \mzp /3$ and $M_S < \mzp$, so that the usual ``forbidden'' setup in which dark matter annihilates into a pair of dark photons is not kinematically accessible.} In the following we will concentrate on accelerator-based analysis. All data points in scan-based plots are compatible with dark matter relic density at the $2\sigma$-level and satisfies the BBN constraints.~\cite{Darme:2017glc}

\section{Dark sector and search strategies at accelerators}

\begin{figure}
	\begin{minipage}{0.49\linewidth}
		\includegraphics[width=1.\linewidth]{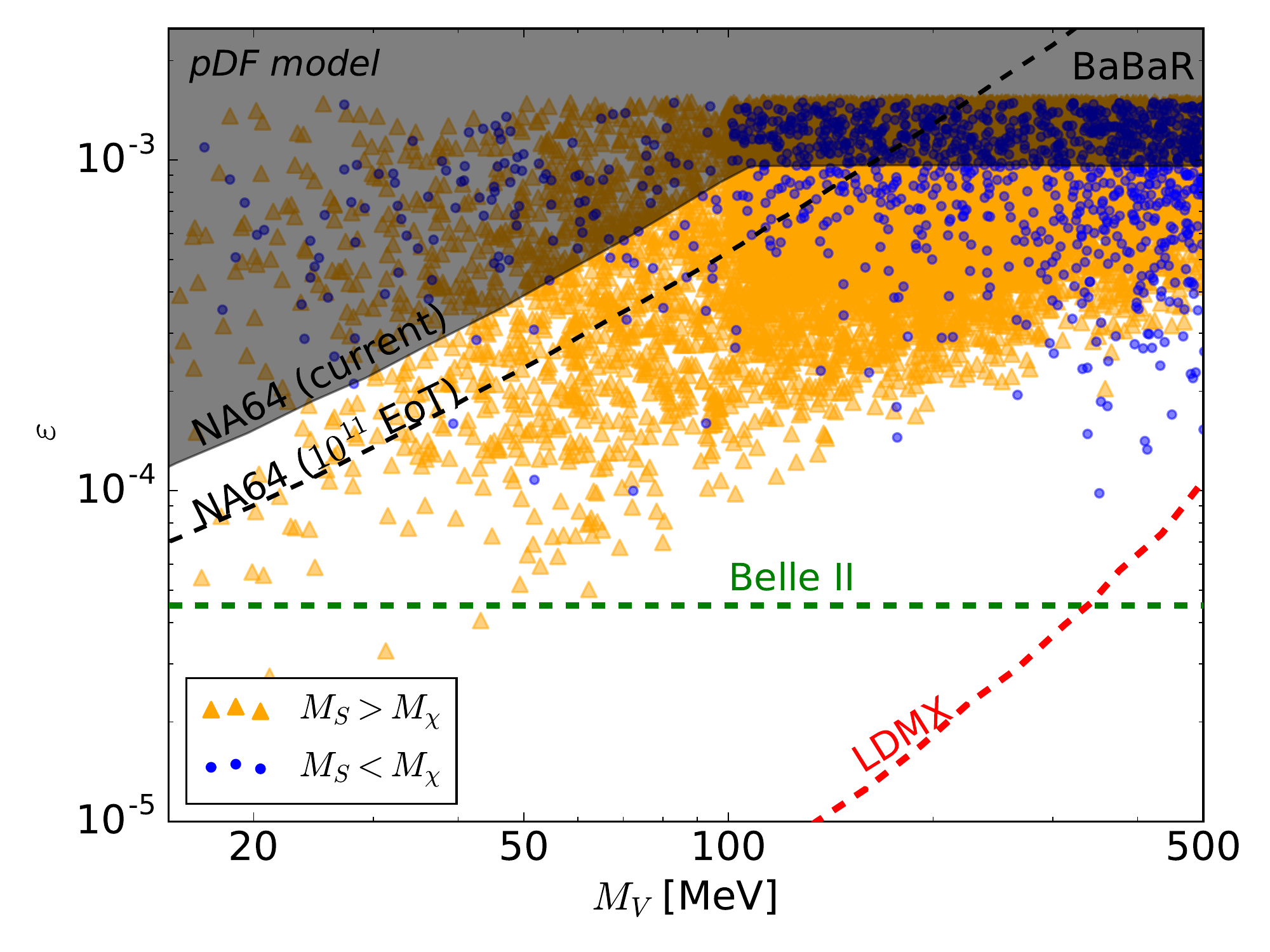}
	\end{minipage}
	\begin{minipage}{0.49\linewidth}
		\vspace{-25pt}
	\includegraphics[width=1.\linewidth]{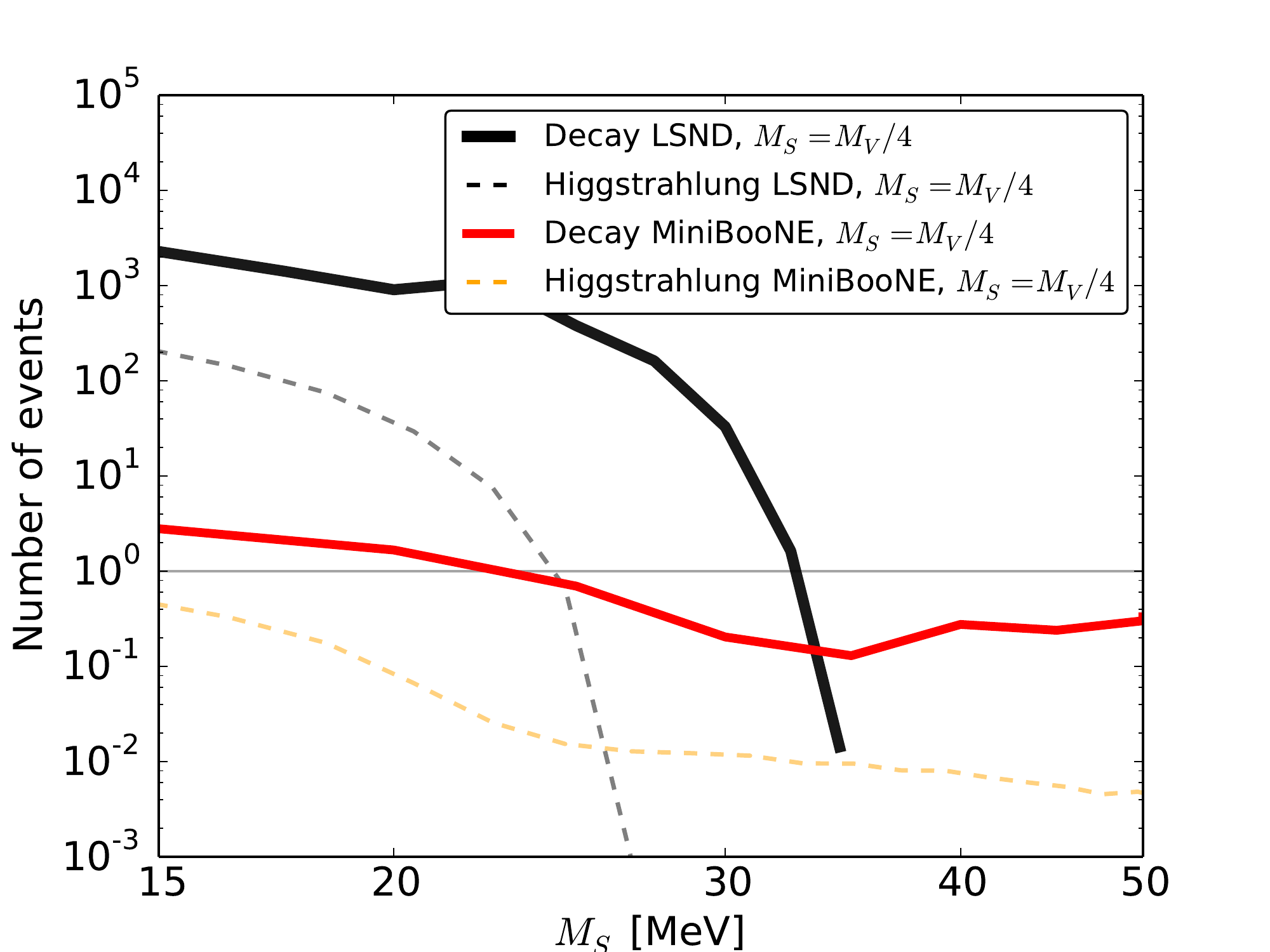}
\end{minipage}
	\hfill
	\caption[]{ (Left) Constraints on the dark photon mass $\mzp$ and on the kinetic mixing parameter $\eps$ from NA64, BaBaR, the projected Belle-II reach (50 ab${}^{-1}$) and LDMX reach based on missing energy searches. (Right) Number of events expected at LSND and miniBooNE from dark Higgs boson decay, either produced through dark Higgsstrahlung (thin dashed) or through heavy sector particle decay (thick plain) as a function of the dark Higgs boson mass $\ms$ for $\ms =  \mzp/4$, $m_{\chi}=  \mzp/3 $ (where $m_{\chi}$ is the bare DM Dirac mass), $\eps = 0.001$, $y_{SR}=0, y_{SL}=0.2$ and $\azp = 0.03$.}
	\label{fig:missingE}
\end{figure}

For each fields of the dark sector, dedicated search strategies have been proposed over the last decade. We will present in the following the most relevant ones, focusing eventually on the particular role of the dark Higgs boson.
\vspace{-4pt}
\paragraph{Mediator/dark photon searches} Searches which focus directly on the dark photon are especially interesting in that they are largely model-independent. 
In all of our parameter space, the dark photon decays invisibly and the strongest bounds come from
missing energy signature in BaBar~\cite{Lees:2017lec} ($\varepsilon \lesssim  1 \times 10^{-3}$) and NA64~\cite{Banerjee:2017hhz} (as low as $\varepsilon \lesssim  1.5 \times 10^{-4}$ for $\mzp \sim 30$ MeV). In the future, the Belle II collaboration should expand these bounds~\cite{Battaglieri:2017aum}  down to $\varepsilon \lesssim  0.8 \times 10^{-4}$. The sensitivity of these analysis scales  as $\varepsilon^2$ so that these strategies have  bright prospects on the long term. Indeed, as we show in Figure~\ref{fig:missingE} they cover already a large portion of the parameter space compatible with the relic density constraint.

\vspace{-4pt}
\paragraph{Dark matter} Since in most of our parameter space the dark photon decays instantaneously into $\chi_1$ and $\chi_2$ any experiments producing it will generate effectively  a ``dark matter beam''. Fixed target experiments designed originally for neutrinos studies are here particularly well-suited. Indeed, dark photons are abundantly produced either through dark bremsstrahlung or light meson decays in the beam dump, the dark matter particles then propagate through the shielding and  subsequently scatter in the detector. This was first pointed out in~\cite{Batell:2009yf,Essig:2009nc} and a rich literature followed. The main drawback of this kind of analysis is the relatively large neutrino background and the fact that the expected number of events typically scales as $\varepsilon^4 \alpha_D$. Current bounds are mainly competitive against missing energy searches in the low dark matter mass region, where bounds from the LSND experiment~\cite{Aguilar:2001ty} can be applied, due to the impressive accumulated dataset of $10^{22}$ light  $ \pi^0$ mesons.

\vspace{0pt}
\paragraph{Heavy dark sector state}  These states are typically produced simultaneously with dark matter from the dark photon decay channels $V \rightarrow  \chi_1 \chi_2 \ , V \rightarrow \chi_2 \chi_2$ whose branching ratios are determined by the dark gauge current from eq.~\eqref{eq:darkcurrent}. When the splitting between both states is smaller than the mass of other dark sector particles (and in particular of the dark Higgs boson mass, as we will see below) then the only decay channel is through an off-shell dark photon
$\chi_2 \rightarrow \chi_1 V^* \rightarrow \chi_1 e^+ e^-$. Defining the splitting $\Delta_\chi \equiv M_{\chi_2} - M_{\chi_1}$, we have the approximate relation:~\cite{Izaguirre:2017bqb}
\begin{equation}
c \tau_{\chi_2}  \propto 100 \text{ m } \times  \left(  \frac{0.1}{\alpha_D} \right)  \left(  \frac{ 10^{-3}}{\eps}\right )^2 \left( \frac{0.2}{\Delta_\chi}\right)^5 \left( \frac{25 \textrm{ MeV}}{\ms}\right)^5 \left( \frac{\mzp}{ 100 \textrm{ MeV}}\right)^4 \ .
\end{equation}
Hence, in the optimal case large portion of the heavy dark states decay directly in the detector so that the expected of number of events can in theory scales as $\varepsilon^2$. Furthermore, the decay signature is an electron-positron pair which can be distinguished from the neutrino scattering events in experiments with basic track reconstructions capability. This approach typically leads to extremely strong bounds on the kinetic parameter,~\cite{Izaguirre:2017bqb} probing values as low as $\epsilon \sim 10^{-6}$. 


\vspace{-4pt}
\paragraph{Dark Higgs boson} Finally, when the dark Higgs boson is light enough not to decay into a pair of DM particle  ($M_S < 2 M_{\chi_1}$), its decay mainly  proceeds through a loop-induced $\varepsilon^2$-suppressed diagram. This leads to an extremely long lifetime of typical order hundreds of seconds  below the di-muon threshold:
\begin{equation}
\tau_S \propto 10 \textrm{ s} \times \left( \frac{\aem}{q_S^2  \azp}\right) \left(  \frac{ 10^{-3}}{\eps}\right)^4 \left( \frac{50 \textrm{ MeV}}{\ms}\right) \left( \frac{\mzp}{ 100 \textrm{ MeV}}\right)^2 \ .
\end{equation}
In principle dark Higgs bosons produced through dark-Higgsstrahlung and decaying into a electron-positron pair can be searched for in neutrino experiments, with  signature similar to the one of a heavy state decay.~\cite{Batell:2009di} However, the very long lifetime implies that the expected number of events scales as $\varepsilon^6 \alpha_D^2$ and current missing energy bounds outperformed such searches in all points of the parameter space realising the correct relic density.~\cite{Darme:2017glc}  The presence of the dark Higgs boson leads therefore to ``blind spots'' in the heavy dark sector decay signatures. Indeed, when the splitting between both $\chi_1$ and $\chi_2$ is large enough, $\Delta_\chi = M_{\chi_2} -  M_{\chi_1} > M_S $ (and provided that $y_{SL}$ is not exactly equal to $y_{SR}$) then the heavy state will decay instantaneously by emitting a dark Higgs. As shown in Figure~\ref{fig:missingE}, this typically increases by an order of magnitude the expected number of events from the dark Higgs boson decay signature described above. On the other hand, the heavy dark state is no longer long-lived so that the bound on $\varepsilon$ is reduced to values typically closer to the DM scattering ones.

\vspace{0.3cm}

To conclude, the presence of an accompanying dark sector is a key ingredient of accelerator-based searches for light thermal dark scenarios, potentially leading to bounds several orders of magnitude larger than the ones deduced from dark matter scattering alone. On the other hand, the high model-dependence of these bounds must be properly accounted for, as demonstrated by the presence of dark Higgs boson-induced ``blind-spots'' in the signature from heavy dark sector state decay.

\section*{Acknowledgements}

LR is supported by the Lancaster-Manchester-Sheffield Consortium for Fundamental Physics under STFC Grant No.\ ST/L000520/1. LD, LR and SR are supported in part by the National Science Centre (NCN) research grant No.~2015-18-A-ST2-00748. The use of the CIS computer cluster at the National Centre for Nuclear Research in Warsaw is gratefully acknowledged.

\section*{References}

\bibliography{LucDarme}






\end{document}